\input{aipcheck}

\documentclass[
    ,final            
  ]
  {aipproc}

\layoutstyle{6x9}

\newcommand{\be}{\begin{equation}}
\newcommand{\ee}{\end{equation}}
\newcommand{\beq}{\begin{eqnarray}}
\newcommand{\eeq}{\end{eqnarray}}

\begin{document}

\title{ Nucleon to $\Delta$ and $\Delta$ form factors in Lattice QCD}


\author{Constantia Alexandrou}{
  address={Department of Physics, University of Cyprus, P.O. Box 20537, 1678 Nicosia, Cyprus and \\
Computation-based Science and Technology Research   Center, 
The Cyprus Institute, P.O. Box 27456, 1645 Nicosia, Cyprus}
}


\begin{abstract}
 We present recent lattice QCD results on the electroweak nucleon to $\Delta$  transition and $\Delta$ form factors using dynamical fermion gauge configurations with a lowest pion mass of about 300 MeV, with special
emphasis in the determination of the sub-dominant  quadrupole  $N\gamma^*\rightarrow \Delta$ and  $\Delta$  electromagnetic form factors. 
\end{abstract}

\maketitle


\section{Introduction}
There is a number of recent lattice QCD calculations
of the electroweak form factors of the nucleon~\cite{Alexandrou:2010cm}. 
The main focus of this presentation is the evaluation of
the electroweak nucleon (N) to $\Delta$ and $\Delta$ form factors (FFs).
In $N\gamma^* \rightarrow \Delta$
  the dominant  magnetic dipole FF,   
 $G^*_{M1}$,  is precisely measured and it therefore serves, like in the
case of the electromagnetic (EM)  FFs of the nucleon,
 as a benchmark of the lattice QCD methodology.
The sub-dominant  quadrupole
FFs $G^*_{E2}$ and $G^*_{C2}$ have also been studied extensively,
since  
their value carries information on the deformation
 in the N/$\Delta$ system. In order to calculate them in lattice QCD, one
 applies several improvements  to attain 
good enough accuracy. 
The dominant
axial N to $\Delta$ form factors $C_5^A$ and $C_6^A$ can also be  calculated within lattice QCD. They correspond to the nucleon
axial,  $G_A$, and
nucleon induced pseudo-scalar, $G_p$, FFs and
 provide important input for phenomenological
models  and   chiral effective theories.
Furthermore, the evaluation of
the $\pi N\Delta$ coupling enables us to check the
validity of PCAC   and the associated non-diagonal Goldberger-Treiman relation.

The FFs of the $\Delta$
 are difficult to measure experimentally due to the $\Delta$ short lifetime. 
Only the magnetic moment of the $\Delta$ is measured albeit with large uncertainty.
Thus lattice QCD can provide a valuable input on  the $\Delta$  FFs  enabling
for instance the determination of
the $\Delta$ charge distribution in the infinite momentum frame~\cite{Alexandrou:2009nj}.  
 Due to 
the 3/2-spin structure of the $\Delta$,  there are two  Goldberger-Treiman relations, which can be examined by calculating, besides the axial $\Delta$ FFs the pseudo-scalar ones.
Having the EM and axial FFs for the N/$\Delta$ system enables
a combined chiral fit to determine the coupling constants that enter
in chiral effective models.

Recently there has been a lot of progress in the calculation of hadron masses
and nucleon structure using dynamical lattice QCD simulations~\cite{Jansen:2008vs}.
 The computational cost of these simulations can be
parameterized as a function of the lattice spacing $a$, the lattice spatial extent $L$ and the pion mass, $m_\pi$, as
$
\small{
 C_{\rm sim}\propto \left ( \frac{300 {\rm MeV}}{m_\pi}\right)^{\ {c_m}}\left( \frac{L}{2 {\rm fm}} \right)^{{c_L}}\left( \frac{0.1 {\rm fm}}{a} \right)^{{c_a}} }
$.
The coefficients ${ c_m,\,c_L}$ and ${ c_a}$ depend on the type of discretized action.
Based on current simulations the cost at
the physical point is estimated of ${\cal O}(1)$~Teraflop$\cdot$year.

\begin{figure}[h!]\vspace*{-0.5cm}
\begin{minipage}{0.4\linewidth}\vspace*{-0.3cm}
\hspace*{-0.8cm}{\includegraphics[width=1.15\linewidth]{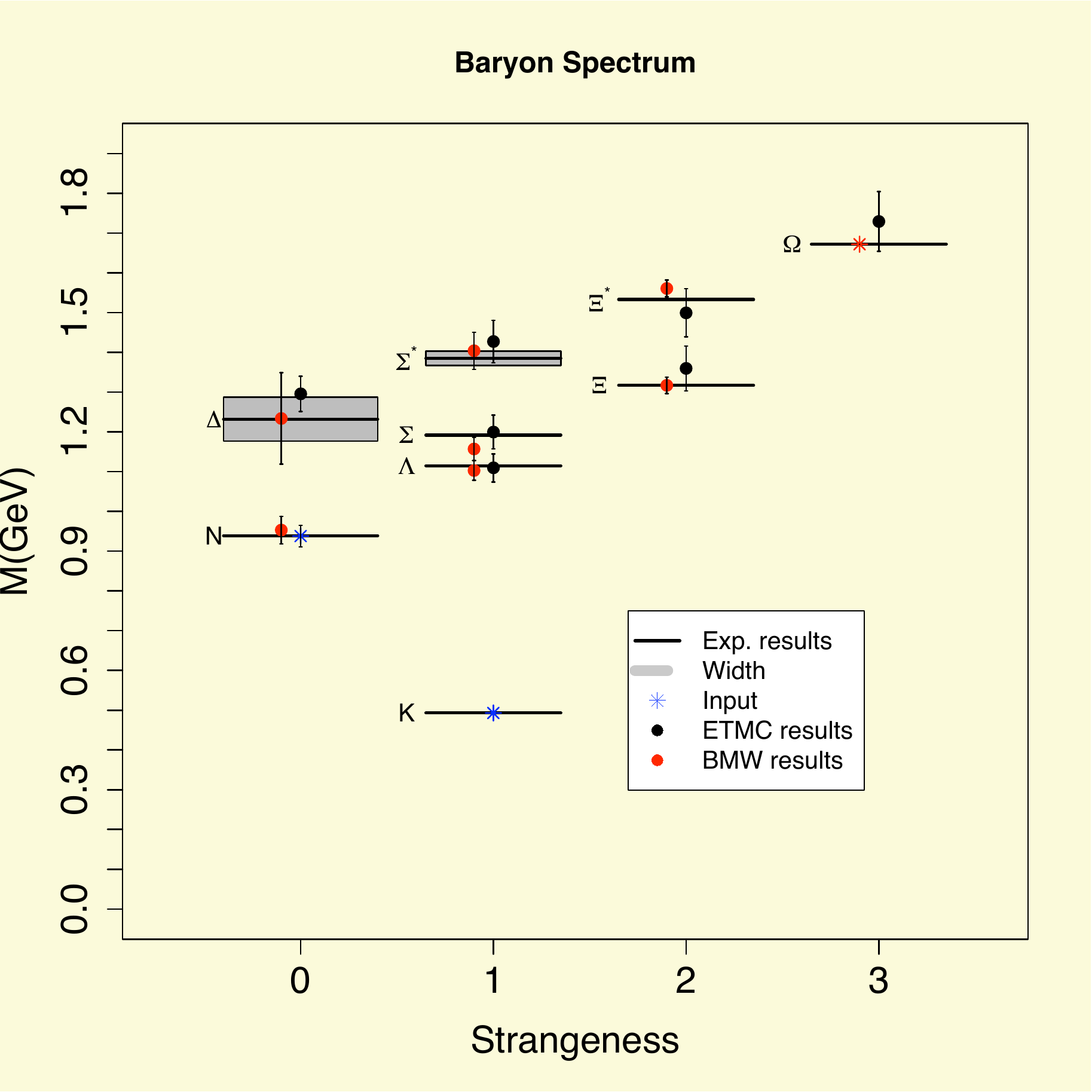}}
\end{minipage}
\begin{minipage}{0.49\linewidth}\vspace*{0.cm}
\includegraphics[width=\linewidth,height=1.\linewidth]{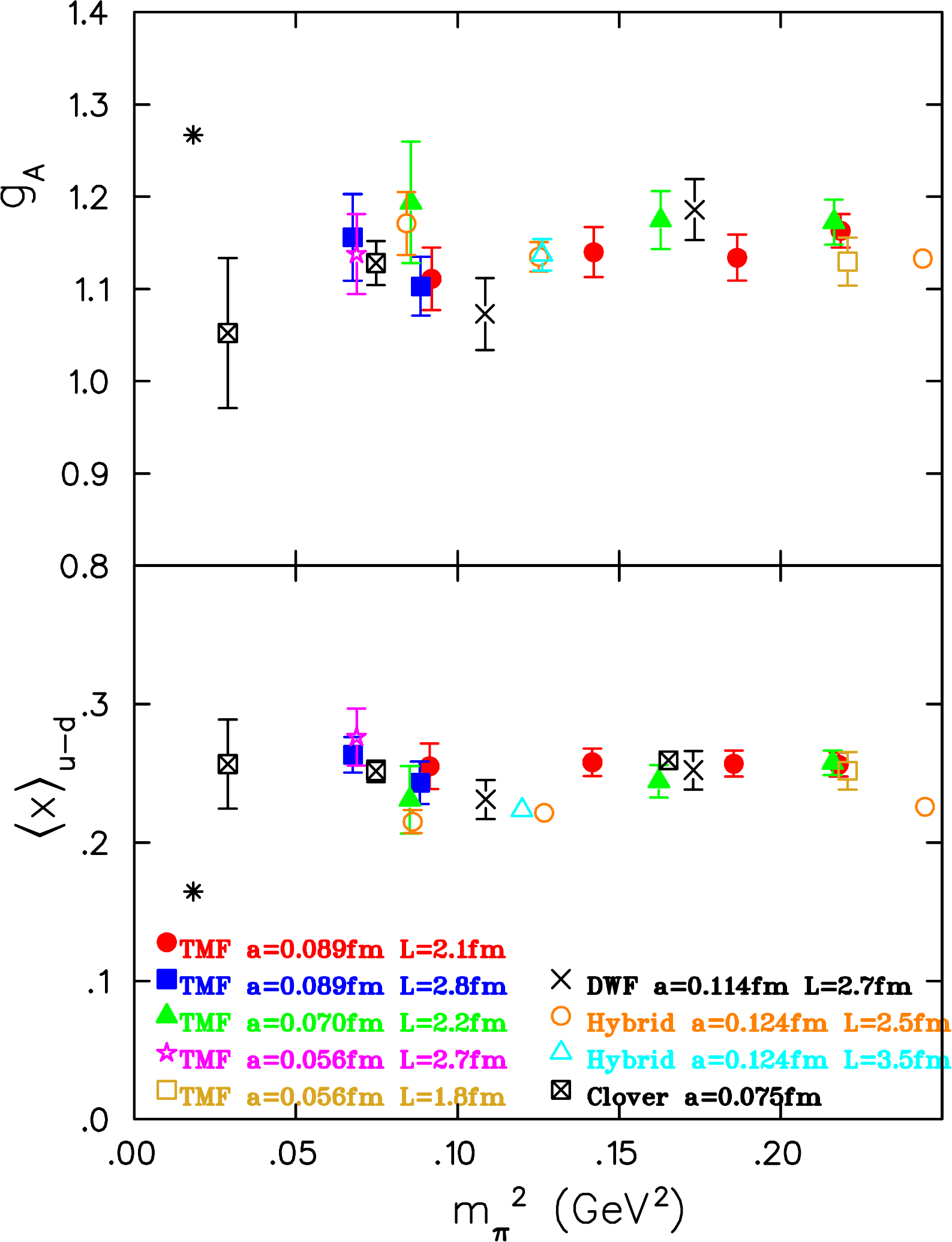}
\end{minipage}
\vspace*{-18cm}
\caption{Left: The low-lying baryon spectrum computed
using $N_f=2+1$ Clover~\cite{Durr:2008zz} and  $N_f=2$ twisted mass fermions (TMF)~\cite{Alexandrou:2009qu}; Upper right:
The nucleon axial charge~\cite{Alexandrou:2010hf}; Lower right: The nucleon isovector moment of the unpolarized quark distribution~\cite{Alexandrou:2011nr}. The physical point is shown by the asterisk.}\label{fig:GFFs}
\end{figure}

In Fig.~\ref{fig:GFFs} we show recent results on the low-lying  baryon spectrum obtained by the  BMW Collaboration using $N_F=2+1$ Clover fermions
and the ETM Collaboration using $N_F=2$ twisted mass fermions.
Both collaborations used 3 lattice spacings
to extrapolate the results to the continuum limit.
One observes that the results using different discretization schemes
are in agreement and that both reproduce the experimental values. This is a  significant validation of
lattice QCD techniques.

The electric 
  ${G_E(q^2)} $ and magnetic 
     ${G_M(q^2)}$  Sachs form factors (FFs)
as well as the  axial-vector FFs  $G_A(q^2)$ and $ G_p(q^2) $ of the nucleon have also been studied by several lattice groups using dynamical 
simulations 
 down to lowest pion mass of typically  $\sim 250$~MeV.  Lattice data are in general agreement, but still show discrepancies with experiment. In Fig.~\ref{fig:GFFs} we compare recent  results from various collaborations on the nucleon
axial charge and isovector moment of the unpolarized quark distribution. Whereas
there is an overall agreement among lattice results the experimental values
are not reproduced (see
Ref.~\cite{Alexandrou:2010cm} for more details).

\section{$N\gamma^*\rightarrow \Delta$  transition form factors}
\noindent

\begin{figure}[h!]\vspace*{-2cm}
\begin{minipage}{0.49\linewidth}
\includegraphics[width=0.9\linewidth,height=0.6\linewidth]{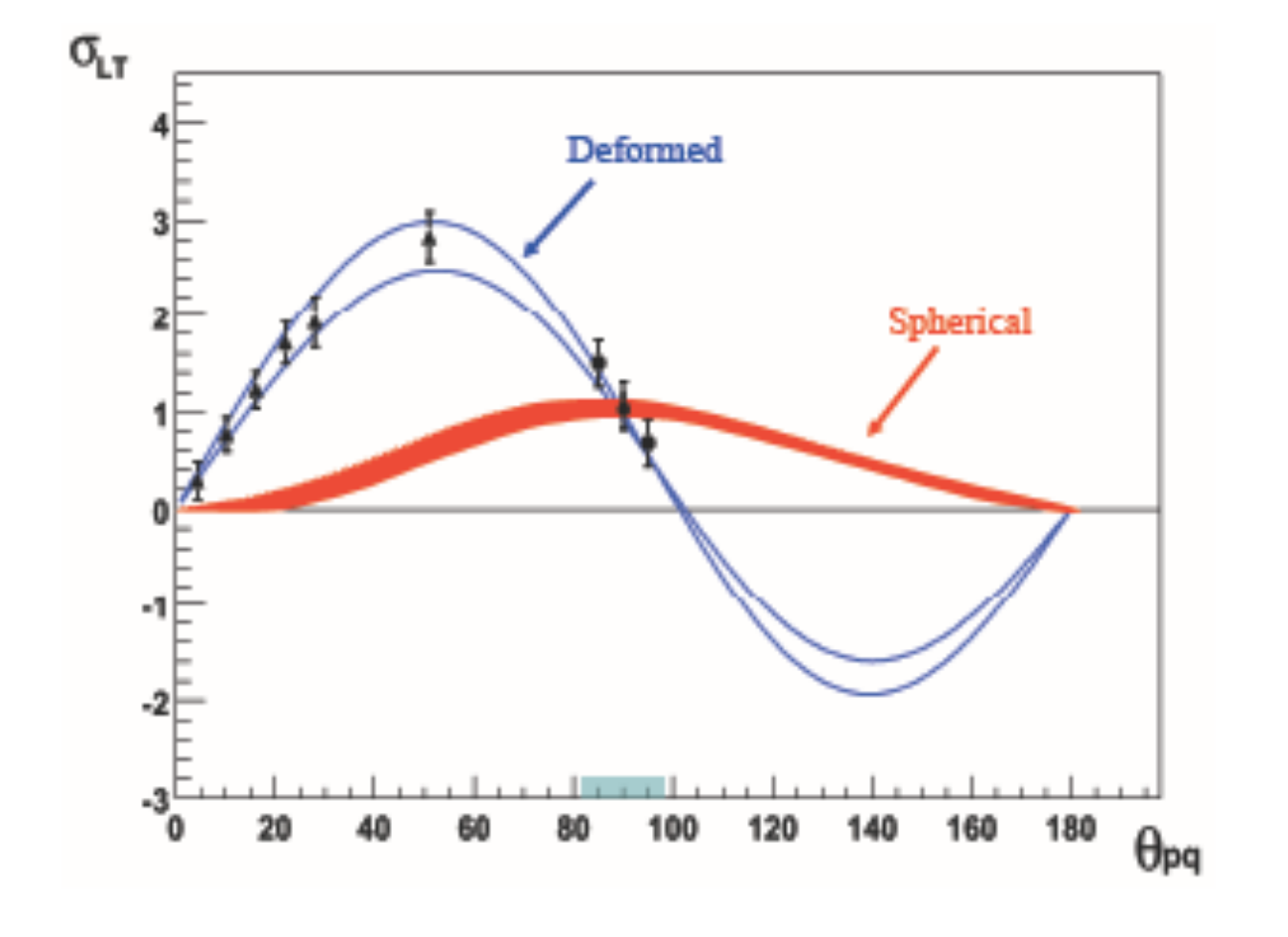}
\end{minipage}\vspace*{-15cm}
\begin{minipage}{0.49\linewidth}
\includegraphics[width=1.\linewidth,height=0.6\linewidth]{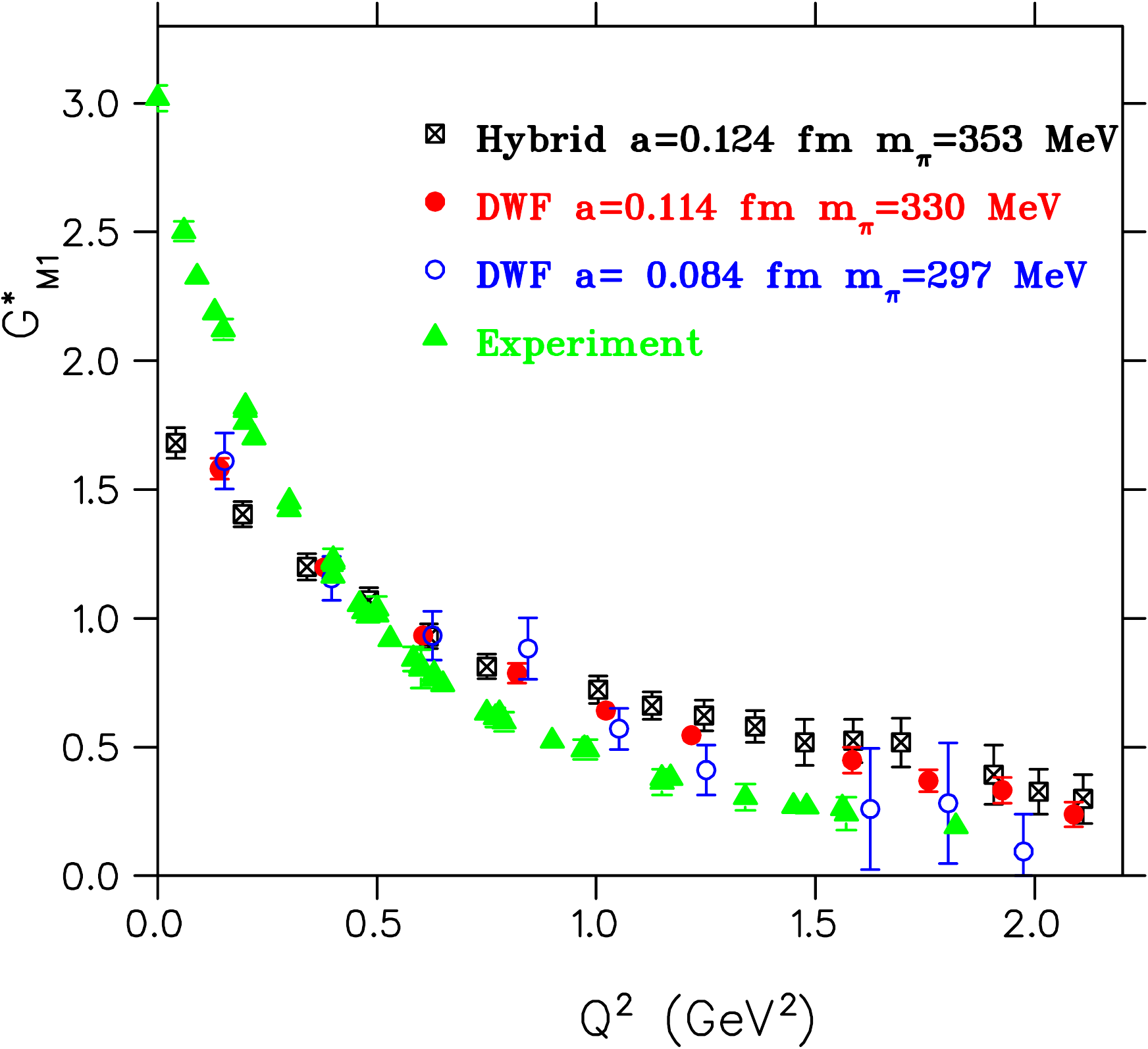}
\end{minipage}
\vspace*{-20cm}
\caption{
Left: The $\sigma_{LT}$ of the $p(e,e^\prime p)\pi^0$ reaction at $Q^2=0.127$~GeV$^2$~\cite{Papanicolas:2003zz}; Right: Experimental and lattice QCD results on the magnetic dipole FF $G^*_{M1}$ as a function of $Q^2=-(p^\prime-p)^2$.}\vspace*{-2cm}
\label{fig:NtoDelta}
\end{figure}

The $N\gamma^*\rightarrow \Delta$ transition is written  in terms of three Sachs FFs:

{\small \be
  \langle\Delta(p',s')\vert j_\mu \vert N(p,s)\rangle =
{\cal A} 
\bar{u}_\sigma (p',s')
  \Biggl[ { G^*_{M1} (q^2)} K_{\sigma\mu}^{M1} + { G^*_{E2}(q^2)}K_{\sigma\mu}^{E2}+{ G^*_{C2}}K_{\sigma\mu}^{C2}\Biggr] u(p,s) 
\label{NtoDelta}
\ee}
\noindent
with ${\small {\cal A}= i\sqrt{\frac{2}{3}} 
\left(\frac{m_\Delta m_N}{E_\Delta({\bf p}^\prime) E_N({\bf p})}\right)^{1/2} }$.
There is a wealth of experimental information on the $N\gamma^*\rightarrow \Delta$ transition~\cite{Bernstein:2007jt}: The dominant magnetic dipole FF, $G^*_{M1}$, is well measured and
the electric $G^*_{E2}$ and Coulomb $G^*_{C2}$ FFs are found to 
be non-zero
 signaling a deformation in the nucleon/$\Delta$-system. 
The deformation is probed via the ratios:
  $ R_{EM}({\rm EMR)} = -\frac{G^*_{E2}(Q^2)}{G^*_{M1}(Q^2)}\>,$ and
  $ R_{SM}({\rm CMR}) = -\frac{\vert\vec{q}\vert}{2m_\Delta}
  \frac{G^*_{C2}(Q^2)}{G^*_{M1}(Q^2)},$
in the rest frame of the $\Delta$.
 As shown in Fig.~\ref{fig:NtoDelta}, precise data strongly ``suggest'' deformation of the N and/or $\Delta$~\cite{Papanicolas:2003zz,Sparveris:2004jn}.
 New data on CMR
at low momentum transfer are currently  being analyzed~\cite{Spaveris}.

\begin{minipage}{0.6\linewidth}\vspace*{0.2cm}
\hspace*{-0.5cm}The lattice evaluation involves the computation of two-
\hspace*{-0.5cm}point and  three-point functions:
   \begin{eqnarray*}     
     G(\vec q, t) &=&\sum_{\vec x_f} \, e^{-i\vec x_f \cdot \vec q}\, 
     {\Gamma^4}\, \langle {J_h(\vec x_f,t_f)}{\overline{J}_{h}(0)} \rangle \\
     G^{\mu\nu}({\Gamma},\vec q, t) &=&\sum_{\vec x_f, \vec x} \, e^{i\vec x \cdot \vec q}\, 
     {\Gamma}\, \langle {J_{h}(\vec x_f,t_f)} {\cal O}^{\mu}(\vec x,t) {\overline{J}_{h}(0)} \rangle
   \end{eqnarray*}
\end{minipage}
\begin{minipage}{0.35\linewidth}\vspace*{-0.8cm}
      \includegraphics[width=\linewidth]{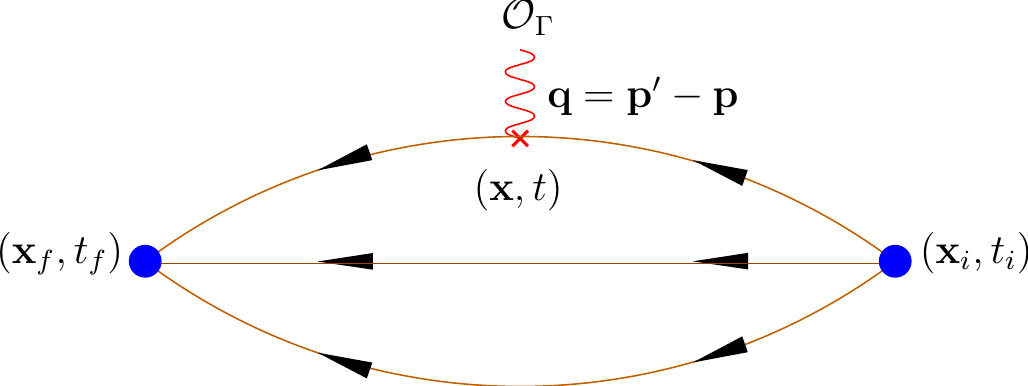}
   \end{minipage}
\noindent
where $\bar{J}_h(x,t)$ is an interpolating field creating a state with the quantum numbers of the baryon $h$ and we have taken $t_i=0$. One computes the
three-point function at various $t$-values of the current insertion ${\cal O}_\Gamma$, which, for
large $t_f$ and $t$ and taking an appropriate ratio with two-point functions, yields the matrix element of Eq.~(\ref{NtoDelta}).
\begin{figure}[h!]\vspace*{-10cm}
\begin{minipage}{0.49\linewidth}
      \includegraphics[width=\linewidth]{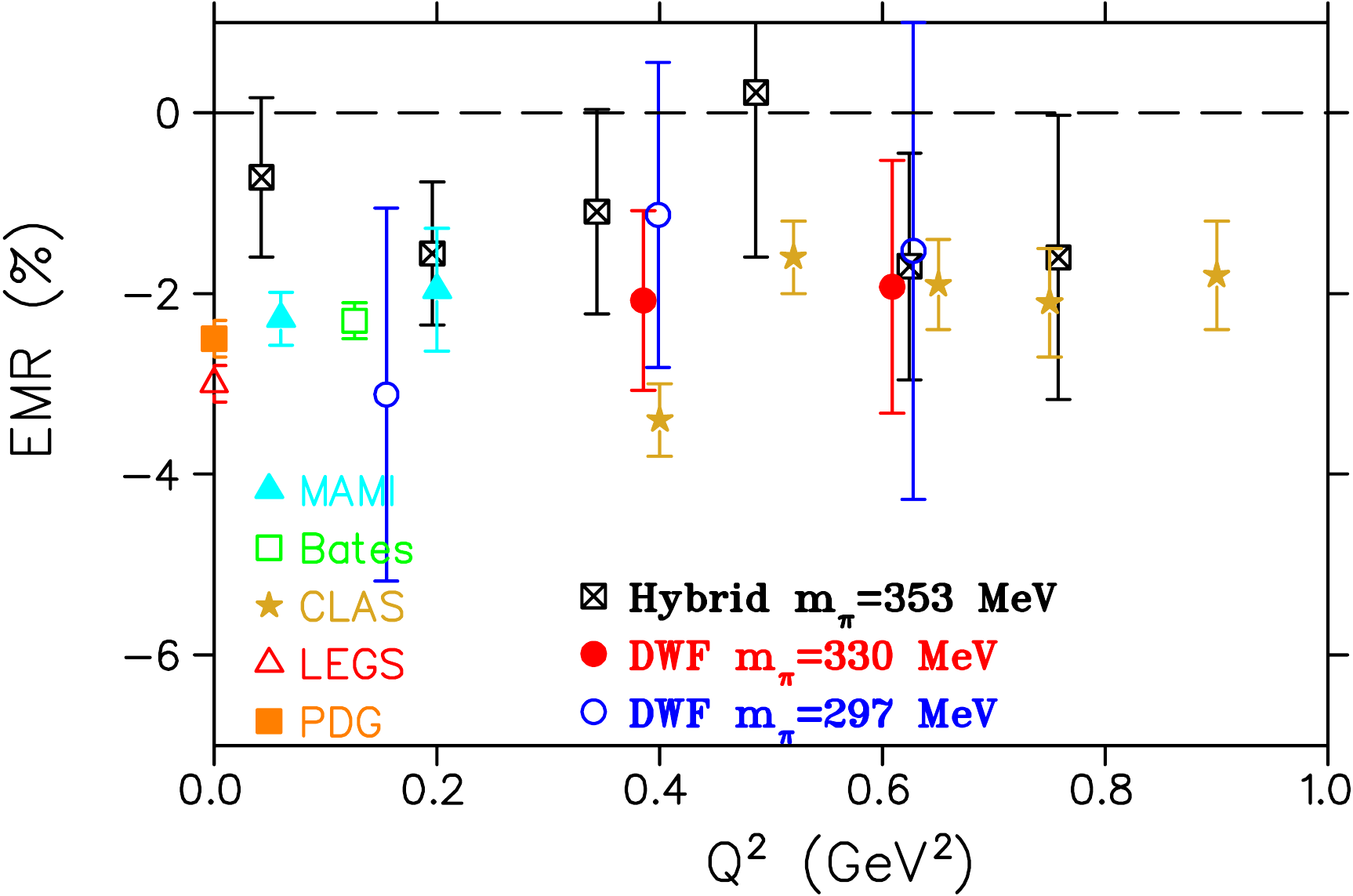}
   \end{minipage}
\begin{minipage}{0.49\linewidth}
      \includegraphics[width=\linewidth]{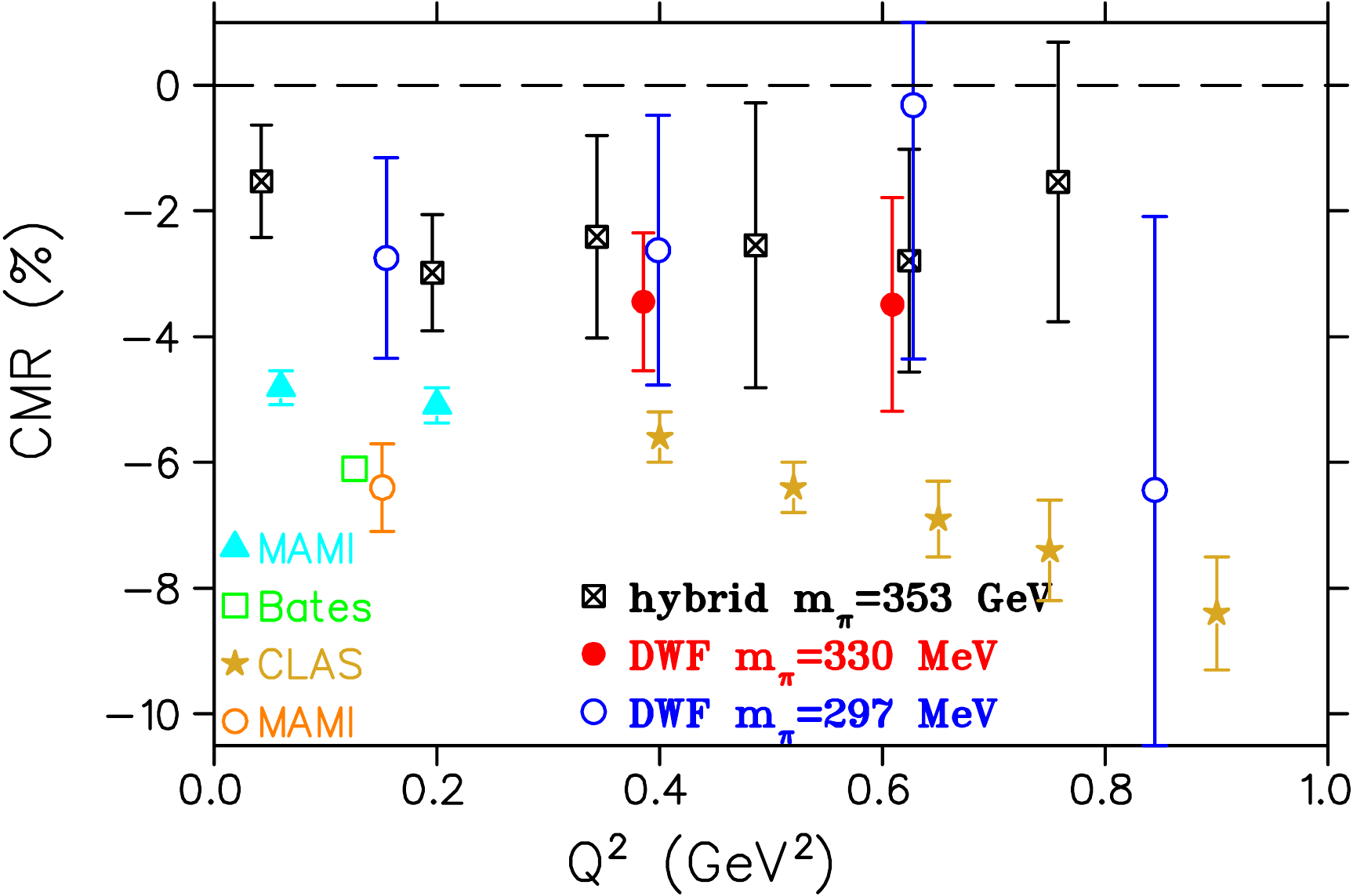}
  \end{minipage}\vspace*{-5cm}
\caption{Results on EMR (left) and CMR (right).}
\label{fig:EMR_CMR}\vspace*{-2cm}
\end{figure}

\noindent
 In order to extract the 
sub-dominant quadrupole FFs one constructs optimized sources to isolate them
from the dominant dipole and uses the coherent sink  technique to increase
statistics.
Recent lattice results are shown in Figs.~\ref{fig:NtoDelta} and \ref{fig:EMR_CMR} 
using a hybrid action of dynamical staggered sea and domain wall valence quarks, as well as $N_F=2+1$ dynamical domain wall fermions (DWF), simulated by the RBC-UKQCD with lowest pion mass of about 300 MeV~\cite{Alexandrou:2010uk}.
The slope of $G^*_{M1}$ at low $Q^2$ remains smaller than what is observed
in experiment underestimating $G^*_{M1}(0)$ i.e. one observes the
same effect as for the nucleon form factors.
Since $G^*_{E2}$ and $G^*_{C2}$ are underestimated at low $Q^2$ like $G^*_{M1}$ taking ratios may remove some of these discrepancies. Indeed the EMR
shown in Fig.~\ref{fig:EMR_CMR} is in better agreement with experiment, whereas
CMR  approaches the  experimental values as the pion mass is lowered. 
Despite the increased statistics  the
errors on the sub-dominant ratios are large
and  to reduce the errors  as $m_\pi$ approaches its physical value  one would need to increase significantly the number 
of statistically independent evaluations.

\section{N -  $\Delta$  axial-vector and pseudo-scalar form factors}

\begin{figure}[h!]
\begin{minipage}{\linewidth}
\includegraphics[width=0.26\linewidth,angle=90]{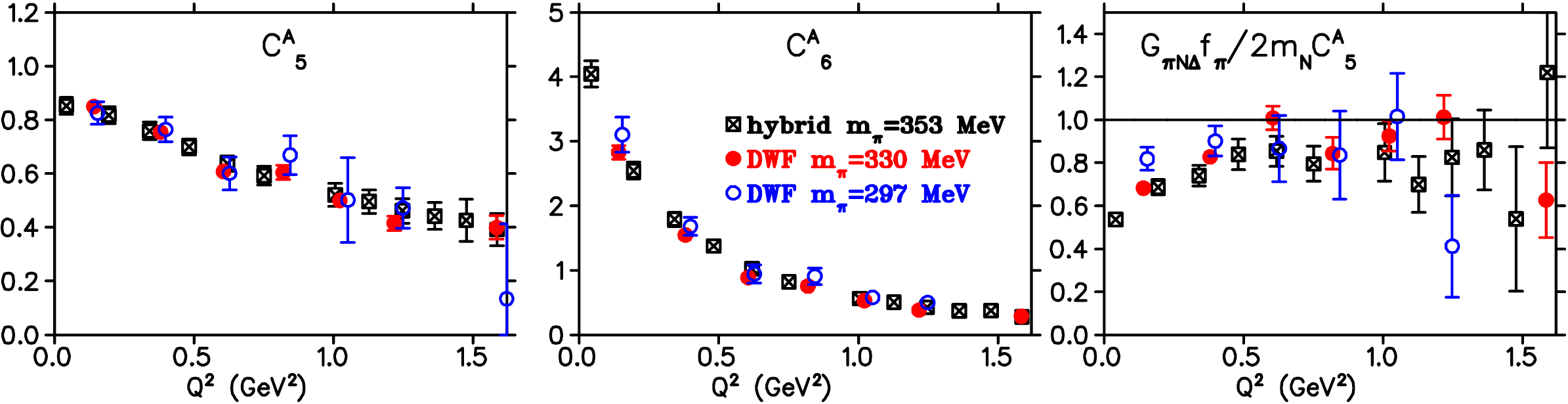}
\end{minipage}
\caption{The dominant axial and pseudo-scalar N to $\Delta$ FFs
 using a hybrid action and DWF. }
\label{fig:axial NDelta}
\end{figure}

The N - $\Delta$ axial-vector matrix element $\langle \Delta(p^{\prime},s^\prime)|A^3_{\mu}|N(p,s)\rangle$ is written as

\be \fontsize{10pt}{12pt}{  
\hspace*{-0.3cm}{\cal A}
\bar{u}^\lambda(p^\prime,s^\prime) 
\Biggl[\left (\frac{{C^A_3(q^2)}}{m_N}\gamma^\nu + \frac{{ C^A_4(q^2)}}{m^2_N}p{^{\prime \nu}}\right)  
\left(g_{\lambda\mu}g_{\rho\nu}-g_{\lambda\rho}g_{\mu\nu}\right)q^\rho 
+{C^A_5(q^2)} g_{\lambda\mu} +\frac{{C^A_6(q^2)}}{m^2_N} q_\lambda q_\mu \Biggr]
u(p,s)}, 
\ee
whereas the N - $\Delta$ matrix element of the pseudo-scalar current is given by
\be {\fontsize{11pt}{12pt} 
 2m_q\langle \Delta(p^\prime,s^\prime)|P^3|N(p,s)\rangle = {\cal A}
\frac{f_\pi m_\pi^2 \>G_{\pi N\Delta}(q^2)}
{m_\pi^2-q^2}
\bar{u}_\nu(p^\prime,s^\prime)\frac{q_\nu}{2m_N} u(p,s)}
\ee
\noindent
Using the axial Ward identity and pion pole dominance
one obtains the non-diagonal Goldberger-Treiman (GT) relation, $G_{\pi N \Delta}(q^2)\>f_\pi = 2m_N C_5^A(q^2)$.
In Fig.~\ref{fig:axial NDelta} we show results on the dominant axial FF $C_5^A$ and on the ratio  $G_{\pi N \Delta}(Q^2) f_\pi /2m_N C_5^A(Q^2)$, which should be unity if the GT relation holds. This ratio approaches unity for $Q^2>0.5$ GeV$^2$~\cite{Alexandrou:2010uk}.

\section{$\Delta$ form factors}
\begin{figure}[h!]
\begin{minipage}{0.33\linewidth}\vspace*{-0.cm}
\hspace*{-0.4cm}{\includegraphics[width=1.1\linewidth,height=0.8\linewidth]{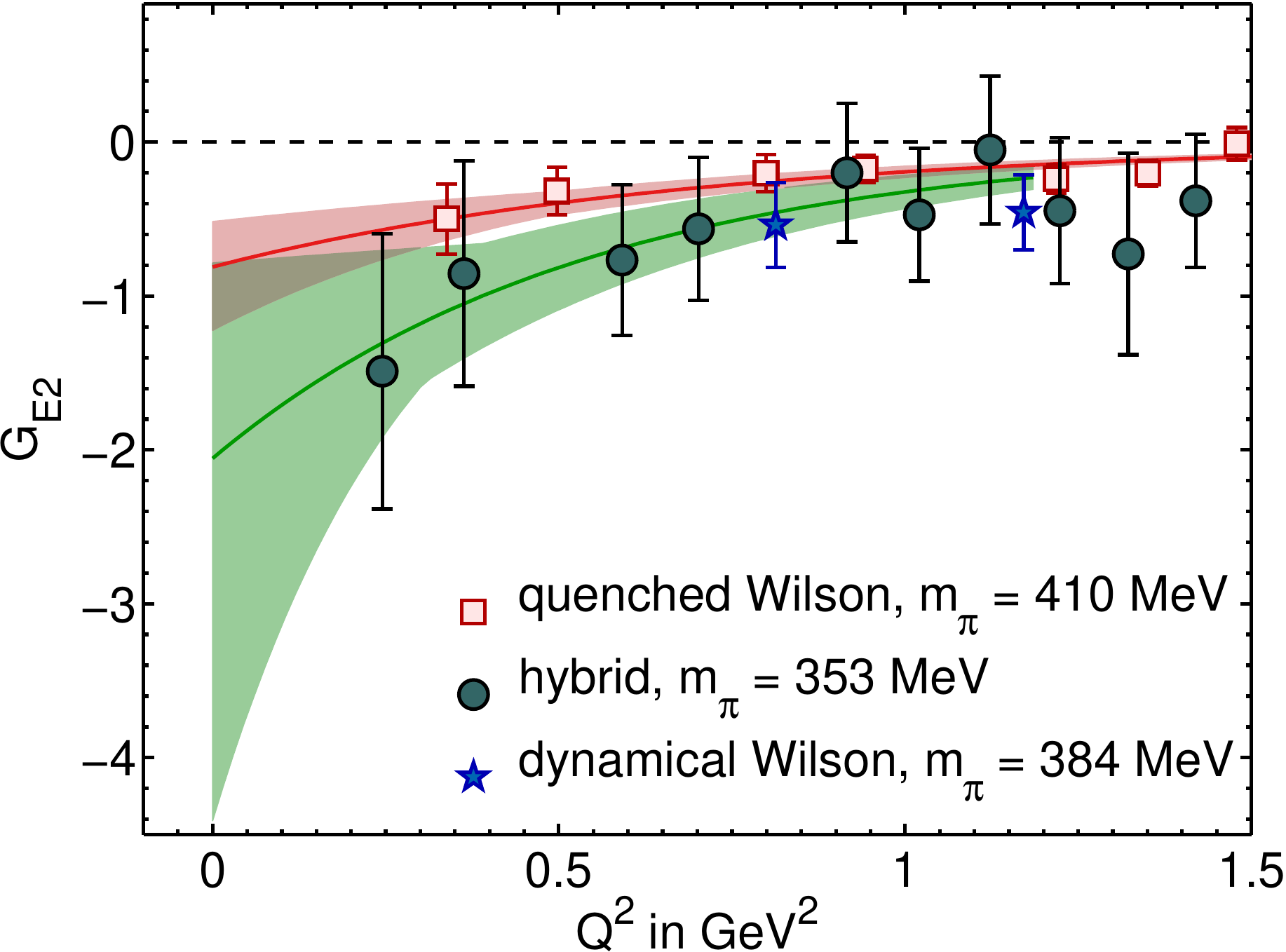}}
\end{minipage}\hfill
\begin{minipage}{0.33\linewidth} \vspace*{0cm}
\hspace*{0.5cm}{\includegraphics[width=0.9\linewidth]{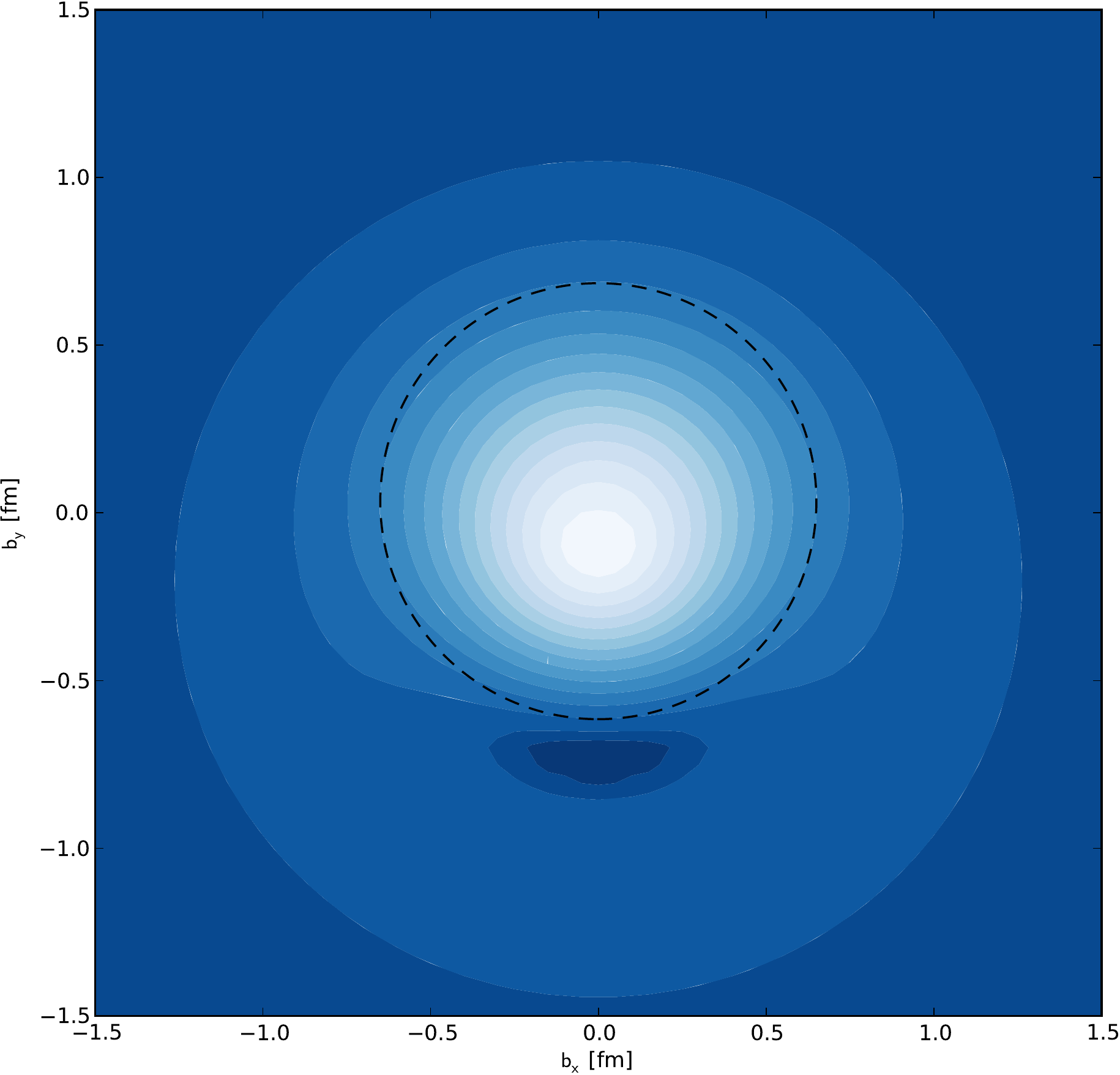}}
\end{minipage}\hfill
\begin{minipage}{0.33\linewidth}\vspace*{0cm}
\hspace*{0.7cm}{\includegraphics[width=0.9\linewidth]{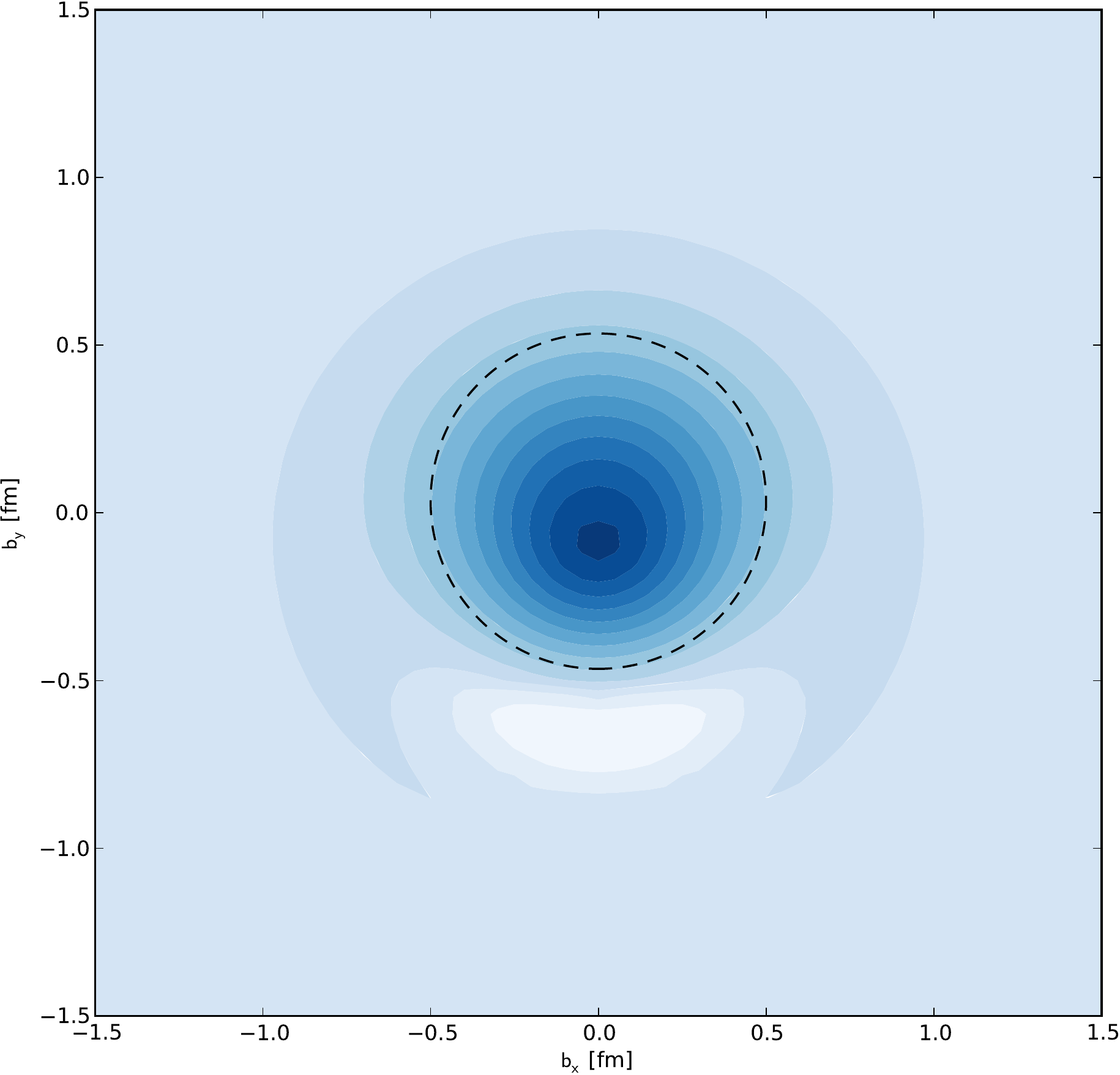}}
\end{minipage}\vspace*{-5cm}
\vspace*{-4cm}\caption{Left: Lattice results on the $\Delta$ electric quadrupole FF; Contours of $\Delta$ (middle) and $\Omega^-$ (right)~\cite{Alexandrou:2010jv}, with 3/2 spin projection along the x-axis. Dark colors denote small values.}
\label{fig:Delta EM}\vspace*{-2cm}
\end{figure}

The $\Delta$ matrix element of the  electromagnetic current $\langle \Delta(p',s^\prime) |j^\mu |\Delta (p,s)\rangle$ is given by

\be \fontsize{9.8pt}{12pt}{  
\hspace*{-0.2cm}-  \bar u_\alpha (p',s^\prime) \left\{  \left[
F_1^\ast(q^2)  g^{\alpha \beta}
+ F_3^\ast(q^2) \frac{q^\alpha q^\beta}{(2 M_\Delta)^2}
\right] \gamma^\mu \right. 
 +\left.\left[ F_2^\ast(q^2)  g^{\alpha \beta}
+ F_4^\ast(q^2) \frac{q^\alpha q^\beta}{(2M_\Delta)^2}\right]
\frac{i \sigma^{\mu\nu} q_\nu}{2 M_\Delta} \, \right\} u_\beta(p,s) \nonumber
}\ee
{\small with e.g. the quadrupole FF given by: 
 $ G_{E2} = \left( F_1^\ast - \tau F_2^\ast \right) - \frac{1}{2} ( 1 + \tau)
\left( F_3^\ast - \tau F_4^\ast \right)$, where $\tau \equiv -q^2 / (4 M_\Delta^2)$}.
Using lattice results on $G_{E2}$ 
one can obtain the transverse charge density of a $\Delta$  in the infinite momentum frame~\cite{Alexandrou:2009hs,Alexandrou:2008bn}.
 This is shown in Fig.~\ref{fig:Delta EM}, where a $\Delta$ with spin 3/2 projection along the x-axis is elongated along the spin axis~\cite{Alexandrou:2009hs}. In the same figure we also show the corresponding charge density of the  $\Omega^-$, which shows a similar  deformation as the $\Delta$~\cite{Alexandrou:2010jv}.

\noindent
The  $\Delta$ matrix elements of the axial-vector current $\langle \Delta(p',s^\prime) |A_\mu^3 |\Delta (p,s)\rangle$ 
  is  given by
 \be \fontsize{9.8pt}{12pt}{   \frac{-1}{2} \bar u_\alpha (p',s^\prime) \left[
g^{\alpha\beta}
\left({g_1(q^2)}\gamma^\mu\gamma^5 
    + {g_3(q^2)} \frac{q^\mu}{2M_\Delta}\gamma^5\right)\right.
+\frac{ q^\alpha q^\beta}{ 4M_\Delta^2}
\left. \left({ h_1(q^2)}\gamma^\mu\gamma^5 
   + { h_3(q^2)} \frac{q^\mu}{2M_\Delta}\gamma^5\right)\right] u_\beta(p,s) 
}\ee
and of the pseudo-scalar current $\langle \Delta(p',s^\prime) |P^3 |\Delta (p,s)\rangle$ by
\be \fontsize{10pt}{12pt}{  -  \bar u_\alpha (p',s^\prime) 
\frac{f_\pi m_\pi^2}{2m_q(m_\pi^2 - q^2)}\left[g^{\alpha\beta}G_{\pi\Delta\Delta}(q^2)\gamma^5 
+\frac{q^\alpha q^\beta}{ 4M_\Delta^2}
H_{\pi\Delta\Delta}(q^2)\gamma^5 \right] u_\beta(p,s) .
}\ee
The $\Delta$ axial charge  is derived from $g_1(0)$~\cite{Alexandrou:2010tj,Alexandrou:2011py}, whereas there are two
  $\pi\Delta\Delta$ pseudo-scalar FFs.

\begin{figure}[h!]\vspace*{-5cm}
\begin{minipage}{\linewidth}
\includegraphics[width=0.26\linewidth,angle=90]{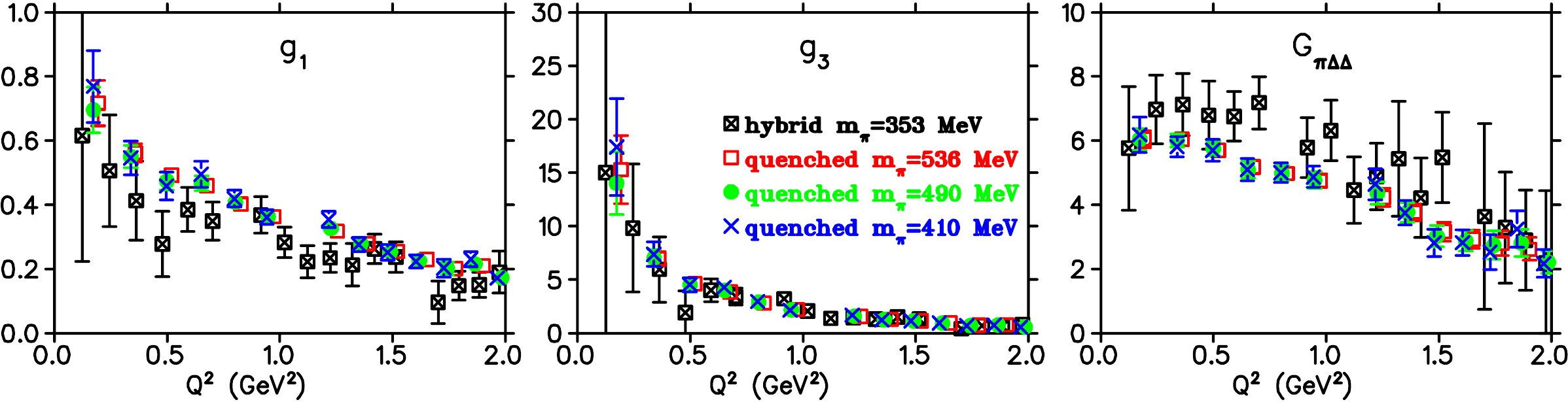}
\end{minipage}\vspace*{-5cm}
\caption{Lattice QCD results on the dominant axial  and pseudo-scalar  $\Delta$ FFs in the quenched theory and using a hybrid action~\cite{Alexandrou:2011py}.}\label{fig:axial Delta} \vspace*{-2cm}
\end{figure}
\noindent
$G_{\pi\Delta\Delta}(0)$  is non-zero and can be identified as the $\pi-\Delta$ coupling. Lattice QCD results on  the dominant axial $\Delta$ FFs $g_1$ and $g_3$ as well as on 
$G_{\pi\Delta\Delta}$  are shown in Fig.~\ref{fig:axial Delta}.
One can derive
  two Goldberger-Treiman relations:
$ f_\pi G_{\pi\Delta\Delta}(q^2) = m_\Delta g_1(q^2)$, and
$
f_\pi H_{\pi\Delta\Delta}(q^2) = m_\Delta h_1(q^2)$, which can be tested using
lattice QCD results.


\section{Conclusions}
We have shown  that lattice QCD successfully reproduces the low-lying baryon
spectrum using different discretization schemes. There is an on-going investigation of nucleon structure by a number of lattice collaborations.
Similar techniques can be  applied to study transitions and resonant properties
and we have applied these methods in the study of the 
$N$ to $\Delta$ electroweak transition FFs as well as  the $\Delta$ FFs. The latter are difficult to measure experimentally and therefore lattice QCD  provides valuable input on these quantities.      Having lattice QCD results on the
$N$-$\Delta$ system one can 
use, for the first time, chiral perturbation theory to 
extract the axial couplings $g_A$, $c_A$ and $g_\Delta$ 
 from a combined chiral fit to the lattice results on the nucleon and $\Delta$ axial charges  and the axial N to $\Delta$ form factor $C_5(0)$~\cite{Alexandrou:2011py}. Applying such a fit to lattice results in the pion mass range from 500 MeV to 300 MeV, still does not reproduce the experimental value of  $g_A$. Current
lattice QCD simulations reaching pion masses below 200 MeV are now
becoming available and  these simulations, combined with a detailed study of lattice systematics~\cite{Dinter:2011sg}, are expected to shed light on the origins
of  the observed discrepancies.

%
\vspace *{-0.2cm}

\begin{theacknowledgments}
{\small I would like to thank my collaborators T. Korzec, G. Koutsou, E. Gregory, J. W. Negele, T. Sato, A. Tsapalis and M. Vanderhaeghen whose contributions
made this
work possible.  I am grateful to G. Koutsou and A. Tsapalis
for providing comments to the manuscript.
This research was partly supported by the Cyprus Research Promotion Foundation (R.P.F) under grant
$\Delta$ IE$\Theta$NH$\Sigma$/$\Sigma$TOXO$\Sigma/$0308/07 and
 by the Research Executive Agency of the European Union under Grant Agreement number PITN-GA-2009-238353 (ITN STRONGnet).
 Domain wall fermion configurations were provided by the RBC-UKQCD collaborations and the forward propagators  by the LHPC and the use of Chroma software~\cite{Edwards:2004sx}.}

\vspace *{-0.1cm}
 
\end{theacknowledgments}



\bibliographystyle{aipproc}   

\vspace *{-0.3cm}

\bibliography{N2Delta_ref}

\IfFileExists{\jobname.bbl}{}
 {\typeout{}
  \typeout{******************************************}
  \typeout{** Please run "bibtex \jobname" to optain}
  \typeout{** the bibliography and then re-run LaTeX}
  \typeout{** twice to fix the references!}
  \typeout{******************************************}
  \typeout{}
 }

\end{document}